\documentclass[prd,twocolumn,floats]{revtex4}
\usepackage{graphicx}

\begin{document}

\title{Trans-Planckian Physics from a Nonlinear Dispersion Relation}

\author{S.E.~Jor\'as}
%\email{joras@if.ufrj.br}
\affiliation{Instituto de F\'{\i}sica, Universidade Fe\-de\-ral do Rio
de Janeiro, Caixa Postal 68528, Rio de Janeiro, RJ 21941-972, Brazil}
\author{G. Marozzi}
%\email{giovanni.marozzi@bo.infn.it}
\affiliation{Dipartimento di Fisica, Universit\`a degli Studi di
Bologna and INFN, via Irnerio 46, I-40126 Bologna, Italy}

\begin{abstract}
We study a particular nonlinear dispersion relation $\omega_p(k_p)$ --- a
series expansion in the physical wavenumber $k_p$ ---
for modeling first-order corrections in the equation of motion of a test
scalar field in a de Sitter spacetime 
from trans-Planckian physics in cosmology.
Using both a numerical approach and a semianalytical one, we show that the WKB
approximation previously adopted in the literature should be used with
caution, since it holds only when the comoving wavenumber $k\gg aH$. 
We determine the amplitude and behavior of the 
corrections on the power spectrum for this test field.
Furthermore, we consider also a more realistic model of inflation, 
the power-law model, using only a numerical approach to determine the 
corrections on the power spectrum.
\end{abstract}

\pacs{98.80.Cq, 98.70.Vc}
\maketitle

%%%%%%%%%%%%%%%%%%%%%%%%%%%%
\section{Introduction}

 Inflationary models provide answers to many problems in standard big bang
cosmology, in particular the origin of density fluctuations and the spectrum
of cosmic microwave background anisotropies. The basis of the whole mechanism
is the stretching of quantum fluctuations generated
at sub-Hubble scales due to the exponential expansion of the spacetime during
inflation. This model, however, has a serious ``problem'': if we consider a
plain scalar-field-driven inflationary model --- say, chaotic
inflation --- the period of inflation lasts so long that the wavelengths of
the fluctuations which at present correspond to cosmological scales were
sub-Planckian at the begining of the inflationary phase.
Therefore, the evolution of fluctuations at such scales is supposed to
follow different rules from those provided by the standard theory of
cosmological perturbations --- which is based on quantum field theory
and general relativity. The set of rules expected to hold above this
energy scale is the so-called trans-Planckian physics (TPP from now
on). The TPP could lead to deviations from standard predictions on the cosmic
microwave background radiation (CMBR), which probes the scales we mentioned
before. The question to be asked is ``whether the predictions
of the standard cosmology are insensitive to effects of TPP''.
This is the precise statement of the 
trans-Planckian problem \cite{rhb99,first}.

In the present paper we adopt, to mimic TPP, 
a series-expansion expression for the
nonlinear dispersion relation $\omega_p$ (Eq.~(\ref{dr}) below), first 
suggested by \cite{w_phys}, to modify the equation of motion of a test scalar
field in a de Sitter spacetime. 
We argue that, in this framework, the WKB approximation is valid only
for $k \gg a H$ (or $k_p \equiv k/a \gg H$ with $k_p$ the physical 
wavenumber and $H$ the Hubble factor). 
As a consequence,  the perturbative approach based on the WKB approximation, 
used in \cite{rhb} to tackle this issue, should be used with caution.

The outline of the paper is the following.
In Sec.~\ref{std} we present the standard approach 
for first-order cosmological perturbations and then we restrict ourselves to 
the 
case of a test scalar field in a de Sitter spacetime. We also introduce the 
WKB approximation in this section. 
In Sec.~\ref{tp} we adopt a particular way to approach TPP: a modification of 
the equation of motion due to a nonlinear dispersion relation. 
The initial conditions for the test scalar field are set in Sec.~\ref{ic}. 
In the following sections we present a numerical calculation, a 3-piecewise 
approximation 
(adopted in Ref.~\cite{rhb}) and a semianalytical approach to solve the 
problem. 
In Sec. \ref{power} we investigate a realistic model of inflation, 
namely power-law inflation, using the same nonlinear dispersion relation.
We then conclude in Sec. \ref{Conc}.

%%%%%%%%%%%%%%%%%%%%%%%%%%%%%%%%%%%%%%%%%
\section{Standard Approach}
\label{std}

The spatially flat Friedmann-Robertson-Walker (FRW)  metric with first-order 
cosmological fluctuations is given by  ($ds^2 = g_{\mu \nu} d x^\mu d x^\nu$):
\begin{eqnarray}
g_{00} &=& a^2 (- 1 - 2 \alpha) \nonumber \\
g_{0i} &=& - \frac{a^2}{2} (\beta_{,i}+B_i) 
\nonumber \\
g_{ij} &=& a^2 \left[ \delta_{ij} (1 -2 \psi)
 + D_{i j} E + \right.\nonumber \\
  && \left. + (\chi_{j,i}+\chi_{i,j}+h_{i j})/2 \right] \,.
\label{metric_general}
\end{eqnarray}
with $D_{ij}=\partial_i \partial_j -1/3 \, \nabla^2 \delta_{ij}$, 
considering the conformal time $\eta$. 
%and having assumed that the 
%tri-curvature vanishes for the sake of simplicity. 
To first-order scalar, vector and tensor perturbations evolve independently.
Vector perturbations can be omitted because they die away kinematically.
The tensor perturbation $h_{i j}$ has only two physical degrees of freedom
(polarization states) $h_{+}$ and $h_{*}$:
\begin{equation}
h_{i j}= h_{+} e^+_{i j}+h_{*} e^*_{i j}
\end{equation}
where $e^+_{i j}$ and $e^*_{i j}$ are the polarization tensors having the 
following properties in the Fourier space:
\begin{equation}
e_{i j}=e_{j i}, \,\,\, k^i e_{i j}=0, \,\,\, e^i_i=0
\end{equation}
\begin{equation}
e^+_{i j} e^{+ i j}=2, \,\,\, e^*_{i j} e^{* i j}=2, \,\,\,
e^+_{i j} e^{* i j}=0
\end{equation}
On expanding in Fourier modes we can define 

\begin{equation}
h_{\lambda} (t, {\vec x}) = \frac{1}{a(\eta)} \frac{1}{(2 \pi)^{3/2}}
\int d^3 k \,
\mu_{t k} (\eta) \, e^{i {\vec k} \cdot {\vec x}}  \,,
\label{quantumFourier}
\end{equation}
with $\lambda=+/*$ and 
where $\mu_{t k}$ will satisfy
\cite{Grish}:
\begin{equation}
\mu_{t k}''+\left(k^2-\frac{a''}{a}\right) \mu_{t k}=0
\label{EqGraviton}
\end{equation}
The scalar sector, in the case of a plain scalar field $\phi(\eta, 
\vec{x})=\phi(\eta)+\varphi(t, \vec{x})$ driven
 inflationary model, can be reduced to the study of a single 
gauge-invariant scalar variable defined by
\begin{equation}
Q=\frac{\mu_s}{a}=\varphi+\frac{\phi'}{{\mathcal H}}\left(\psi+\frac{1}{6}
\nabla^2 E\right)
\label{Mukkhanov}
\end{equation}
where $\mathcal H=\frac{a'}{a}$. This so-called Mukhanov 
variable \cite{Muk} obeys the following equation of motion in Fourier space:
\begin{equation}
\mu_{s k}''+\left(k^2-\frac{z''}{z}\right) \mu_{s k}=0
\label{EqMuk}
\end{equation}
with $z=a\frac{\phi'}{\mathcal H}$.
Considering an adiabatic vacuum state as initial condition for 
those perturbations, all the statistical properties are characterized
 by the two-point correlation function, namely by the 
power spectrum. The dimensionless power spectrum for scalar and tensor 
cosmological fluctuations are given, respectively, by
\begin{equation}
P_Q=\frac{k^3}{2 \pi^2} \left|\frac{\mu_{s k}}{z}\right|^2 \quad,\quad
P_h=\frac{2 k^3}{\pi^2} \left|\frac{\mu_{t k}}{a}\right|^2.
\label{PSCosmo}
\end{equation}
Their dependence on the mode $k$ is defined by the spectral index in the 
following way:
\begin{equation}
n_s-1\equiv \frac{d \ln P_Q }{d \ln k} \quad,\quad
n_t\equiv \frac{d \ln P_h}{d \ln k} 
\label{PSCosmoSI}
\end{equation}
evaluated at a scale $k\ll a H$ when the mode is outside the horizon. 
For $n_s=1$ one has a scale-invariant spectrum for the gauge-invariant 
cosmological scalar fluctuation.

Let us now restrict ourselves to the case of a test scalar field in a de 
Sitter spacetime where $a(\eta)= -1/(H\eta)$. In this case, cosmological 
fluctuations vanish identicaly.

If we expand the test scalar field $\phi(\eta,{\vec x})$ in Fourier
modes
\begin{equation}
\phi(\eta,{\vec x})=\frac{1}{a(\eta)}\frac{1}{(2 \pi)^{3/2}}\int d^3k\,
\mu_k(\eta) \, e^{i {\vec k} \cdot {\vec x}}\quad ,
\label{test_field_expansion}
\end{equation}
then the equation of motion for each mode is given
by
\begin{equation}
\mu_k''+ \Omega^2(\eta) \, \mu_k=0
\label{mu}
\end{equation}
with
\begin{equation}
\Omega^2(\eta) \equiv k^2 - \frac{a''}{a}
\label{TF_equation_eta_a}
\end{equation}
which becomes for a de Sitter spacetime
\begin{equation}
\Omega^2(\eta) = k^2-\frac{2}{\eta^2}\,.
\label{TF_equation_eta}
\end{equation}
We can note that Eq.(\ref{mu}) is the same equation of 
motion as that of a tensor perturbation (\ref{EqGraviton}). 
In this simple case, Eq.~(\ref{mu}) can be exactly solved. 

The two-point correlation function is given by
\begin{equation}
\langle 0| \phi (\eta,{\vec x}) \phi (\eta,\vec{x}+\vec{r}) |0 \rangle
=\int_0^{+\infty} \frac{dk}{k}\frac{\sin kr}{kr} P_\phi (k)
\label{TF_2pointfunction}
\end{equation}
and the power spectrum is
\begin{equation}
P_\phi (k)=\frac{k^3}{2 \pi^2}  \left|\frac{\mu_k}{a}\right|^2.
\label{TF_powerspectrum}
\end{equation}
For superhorizon modes ($k\ll a H$) the spectrum is time independent
and scale invariant, as one can see from the exact solution in the limit 
$\eta\rightarrow 0^-$.
 
Equation (\ref{mu}) can also be interpreted  as a Schr\"odinger equation for a 
stationary wave function with energy $E\equiv \omega^2=k^2$ in an 
effective potential $V_{\rm eff}(\eta) \equiv a''/a$ which is a 
function of the ``spatial'' variable
$\eta$. We might be tempted to solve this equation using WKB approximation,
just as it is usually done in basic quantum mechanics
(QM) \cite{merz}. In this approximation the stationary solution of
Eq.~(\ref{mu}) is given by
\begin{eqnarray}
\mu_k(\eta) &=& \frac{c_+}{\Omega^{1/2}}
 \exp\left[+i \int^\eta \Omega(\eta') \, d\eta'\right]
+\nonumber \\
 &+&  \frac{c_-}{\Omega^{1/2}} \exp\left[-i \int^\eta \Omega(\eta') \,
d\eta' \right]
\label{wkb}
\end{eqnarray}
as long as the WKB parameter $W$ is much smaller than 1:
\begin{equation}
W\equiv \left|\frac{1}{\Omega^2}\left[\frac{3}{4}
\left(\frac{\Omega'}{\Omega}\right)^2 -
\frac{1}{2} \frac{\Omega''}{\Omega}\right]\right| \ll 1 \qquad .
\label{wkb_eq}
\end{equation}
\begin{figure}
\includegraphics[angle=-90,width=0.5\textwidth]{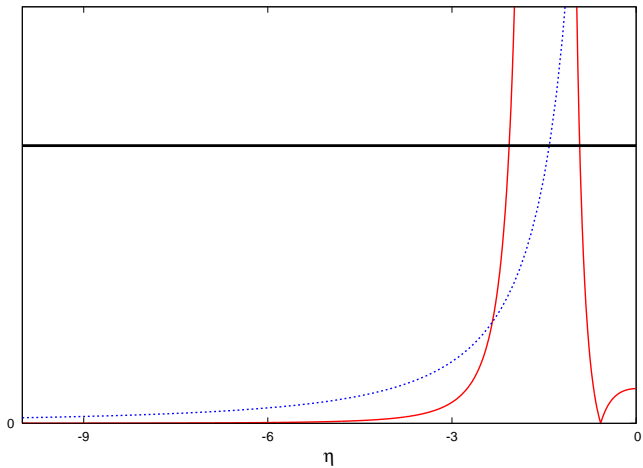}
\caption{Total energy (full straight horizontal line), effective potential 
(dashed blue curve), and WKB parameter (full red curve) as a function of the 
conformal time $\eta$ for a linear dispersion relation. We have used $k=M_{pl}$
and $\eta$ is in units of $1/M_{pl}$.}
\label{qm1}
\end{figure}

Figure \ref{qm1} shows the behavior of $W$ as a function of the conformal 
time $\eta$ in a de Sitter case. 
According to this plot, the WKB approximation holds for 
$\eta \rightarrow -\infty$ ($k \gg a H$, subhorizon scales) but not for 
values close to zero ($k \ll a H$, superhorizon scales), where it equals 
$0.125$ (i.e, {\it not} much smaller than $1$)  \cite{wkb}. 
It fails exactly where it  is supposed to: near the classical turning point
($\eta_{\rm tp}\equiv -\sqrt{2}/k$) and where the effective potential is
too steep ($\eta\rightarrow 0^-$). For the subhorizon scales, 
where $E\gg V_{\rm eff}$, the solution of
Eqs.~(\ref{mu},\ref{TF_equation_eta}) is a plane wave in conformal time 
with comoving frequency $\Omega \simeq k$, Eq.(\ref{wkb}), as expected.

%%%%%%%%%%%%%%%%%%%%%%%
\section{Trans-Planckian behavior: non-linear dispersion relation}
\label{tp}

A modification of the linear dispersion relation
(d.r. from now on) was proposed by Unruh \cite{unruh} 
for describing high-energy Physics in the black-hole
radiation emission. He was inspired by sound waves, for which a linear d.r.
 ceases to be valid when the wavelength gets closer to or
smaller than the lattice spacing. Jacobson and Corley \cite{cj} (see also
\cite{first}) also
proposed nonlinear terms in the d.r. that could be justified by the
inclusion of higher-order derivatives in the Lagrangian. Other changes can
also be introduced by arguing that the spacetime symmetries migh not survive
at high energies \cite{kg}.

Following this approach to mimic TPP  one writes the comoving 
frequency $\omega$ as $\omega=a(\eta) \omega_p(k_p)$ --- 
where $\omega_p$
is the physical frequency --- and assumes that $\omega_p$ is a nonlinear
function of $k_p$ which differs from the standard (linear) one only for
physical wavelengths closer to or smaller than the Planck
scale. Note that this replacement would be
innocuous if the dispersion relation was linear in $k_p$.

\begin{figure}
\includegraphics[angle=-90,width=0.5\textwidth]{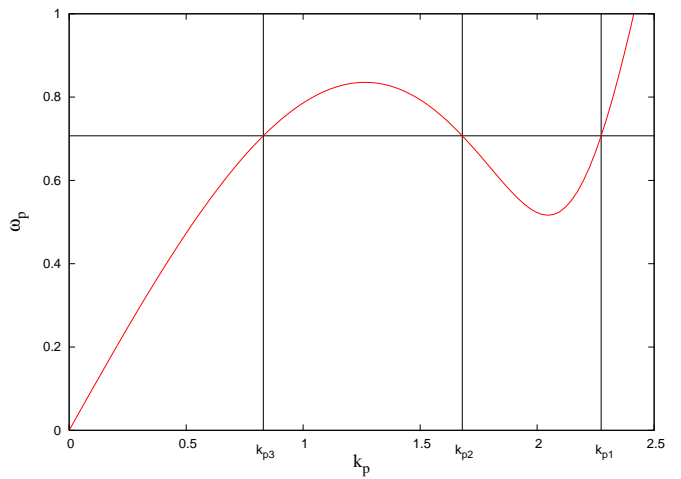}
\caption{Nonlinear dispersion relation as a function of
the physical wavenumber $k_p$ for a case with three different solutions
($k_{p 1}$, $k_{p 2}$ and $k_{p 3}$) of the turning-point equation 
$\omega_p^2=2 H^2$. $H=0.5 M_{pl}$ and $k_p$ is in units of $M_{pl}$.}
\label{nldr}
\end{figure}

In this paper we focus on the d.r.
\begin{equation}
\omega_p^2 (k_p) = k_p^2 - \alpha k_p^4 + \beta k_p^6\quad,
\label{dr}
\end{equation}
with $\alpha>0$ and $\beta>0$, proposed in Ref.~\cite{w_phys}. 
See Fig.~\ref{nldr} for 
a sketch of this function. The above expression can be seen as a mere series 
expansion, but it is also found in solid state physics and describes 
the {\it rotons}~\cite{abrik}. There has been suggestions \cite{myers} for 
including a 3-order term in the nonlinear d.r. above, coming from an 
effective-field-theory approach. Since such odd-power terms violate CP, 
we do not consider them here. 

\begin{figure}
\includegraphics[angle=-90,width=0.5\textwidth]{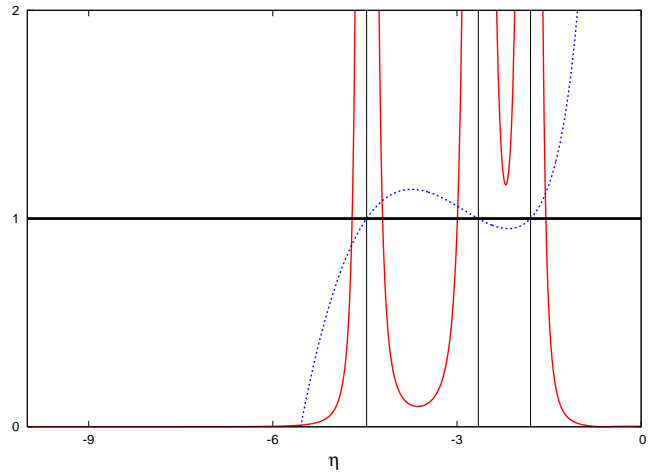}
\caption{Total energy (full straight horizontal line), effective potential 
(dashed blu curve), and WKB parameter (full red curve) as a function of the 
conformal time $\eta$ for the nonlinear dispersion relation Eq.~(\ref{dr}). 
We have used $\bar \alpha = 0.1325$, $\bar \beta=0.004375$ and $H=0.5 M_{pl}$ 
so that 
there are 3 turning points (indicated by the vertical lines). The WKB curve 
was multiplied by $2 \cdot 10^{-2}$ so its behavior in the region close to 
the turning points can be clearly seen. Note, however, that the 
WKB approximation 
is valid only in the region $\eta\rightarrow -\infty$ since 
$W(\eta\rightarrow 0^-)= 0.125$. We have used $k=M_{pl}$
and $\eta$ is in units of $1/M_{pl}$.}
\label{qm2}
\end{figure}

Following the discussion in Section~\ref{std}, we plot in Fig.~\ref{qm2} the
analogous quantities to $W$ (\ref{wkb_eq}), the total energy $E\equiv k^2$,
and the effective potential, the latter being given by
\begin{equation}
V_{\rm eff} (\eta) \equiv (\bar\alpha k^4)\eta^2 - 
(\bar\beta k^6)\eta^4+\frac{2}{\eta^2},
\label{veff}
\end{equation}
where $\bar\alpha \equiv \alpha H^2$ and $\bar\beta \equiv \beta H^4$ are 
dimensionless quantities.
As in the linear case, the WKB approximation is valid only
when
$\eta \rightarrow -\infty$.

To compare our results with Ref.~\cite{rhb} we consider only a
particular range of values for $\alpha$ and $\beta$, necessary
for a positive nonlinear dispersion relation with three real
distinct solutions for the classical turning-points equation
$\omega_p^2=2 H^2$ (see, also, \cite{rhb2}). 
The request to have three real turning-points and a nonlinear 
$\omega_p$ that differs from the standard linear one only for
physical wavelengths closer to or smaller than the Planck
scale has as consequence that $H$ should be comparable with $M_{pl}$. In
fact, everywhere in the paper we consider $H=0.5 M_{pl}$.

Defining $z \equiv 3 \bar{\beta}/\bar{\alpha}^2$ one 
obtains the following ranges for $z$ and $\bar\alpha$:
\begin{equation}
3/4<z<1 \quad,\quad g(z)<\bar \alpha<f(z)
\label{range}
\end{equation}
with
\begin{equation}
g(z)\equiv \frac{1}{3z^2}\left[\frac{3z}{2}-1-(1-z)^{3/2}\right]
\label{grange}
\end{equation}
\begin{equation}
f(z) \equiv \frac{1}{3z^2}\left[\frac{3z}{2}-1+(1-z)^{3/2}\right]
\label{frange}
\end{equation}
where $z>3/4$ is the physical condition to have a positive nonlinear d.r..
The problem of finding the power spectrum from this particular nonlinear
dispersion relation has been tackled in Ref.~\cite{rhb}. In that paper the 
authors 
introduced an approximate piecewise form of the nonlinear dispersion relation
(see Sec.\ref{LA}) 
and apply the WKB approximation for $k_p$ such that $\omega_p^2 > 2 H^2$.
They argue that the amplitude of the effects of the nonlinearity is 
proportional to $\Delta \equiv (k_{p 1}-k_{p 2})/k_{p 1}$  where $k_{p 1}$ and 
$k_{p 2}$ are,
respectively, the first and the second turning points from the right of
Fig.~\ref{nldr}. Still according to Ref.~\cite{rhb}, $\Delta$ measures 
the ``time" spent in the region where the WKB approximation is {\it not} 
satisfied, but it actually is simply a measure of the distance between the 
two largest turning points --- $k_{p 1}$ and $k_{p 2}$ in Fig.~\ref{nldr}. 
It is clearly seen in Fig.~\ref{qm2} that $\Delta$ is much smaller that such 
time interval.
 
We show below the full result (i.e, the exact solution within the 
three-piecewise approximation) and two other different
approaches --- one numerical and the other semianalytical --- for evaluating
the nonlinear effects on the power spectrum.

But before proceeding to proving the claims above, we shall first determine
the initial conditions we assume.

%%%%%%%%%%%%%%%%%%%%%%%%%%%%%%%%%%%%%%%%%%
\section{The equation of motion: initial conditions}
\label{ic}

The equation of motion for the mode function of
a test scalar field, following the approach just introduced to 
consider the TPP effects, is given by
\begin{equation}
\mu_k''+ \left(a^2 \omega_p^2-\frac{a''}{a}\right) \mu_k=0\, 
\label{TF_equation_eta2}
\end{equation}
with $\omega_p$ given by Eq.(\ref{dr}).

For $k \gg a H$ the WKB approximation is valid (see Fig.~\ref{qm2}), 
the term $a^2 \omega_p^2 = \omega^2$
dominates and the mode function is given, chosing the adiabatic vacuum 
(see \cite{rhb}), by:
\begin{equation}
\mu_{wkb}(k, \eta) = \frac{1}{\sqrt{2 \omega(k, \eta)}}\exp
\left[-i \int_{\eta_i}^{\eta} \omega(k, \tau) \, d\tau \right] \,. 
\label{WKB_Solution}
\end{equation}
To proceed with either a numerical or an analytical approach to find the 
solution of
Eq.(\ref{TF_equation_eta2}) we have to find the initial conditions associated 
to our nonlinear dispersion relation in a de Sitter spacetime. 
We consider a initial time
$\eta_i$ and a fixed value of $k$ (which should be much larger than 
$a(\eta_i) H$).

In this case, for this fixed value $k \gg a H$, the comoving
frequency $\omega$ becomes nearly equal to $\sqrt{\beta}k^3/a^2$
and the WKB vacuum is given by
\begin{eqnarray}
\mu_{wkb}(\eta) &=&
-\frac{1}{\sqrt{2}\beta^{1/4} k^{3/2}} \frac{1}{H \eta} \times \nonumber \\
&& \times \exp\left[ -i
 \beta^{1/2} k^3 \frac{H^2}{3}(\eta^3-\eta^3_i)\right] \quad,
\label{WKB_NLDR}
\end{eqnarray}
its derivative by
\begin{eqnarray}
\mu_{wkb}'(\eta) &=& \left\{
\frac{1}{\sqrt{2}\beta^{1/4}k^{3/2}}\frac{1}{H \eta^2} +i
\frac{1}{\sqrt{2}}\beta^{1/4} k^{3/2}H \eta
\right\} \times \nonumber\\
&&\times \exp\left[ -i \beta^{1/2}
k^3 \frac{H^2}{3}(\eta^3-\eta^3_i)\right]\,,
\label{WKBderivative_NLDR}
\end{eqnarray}
and our initial conditions by
\begin{eqnarray}
\mu_{wkb}(\eta_i) &=& -\frac{1}{\sqrt{2}\beta^{1/4} k^{3/2}}\frac{1}{H \eta_i}
\label{WKB_NLDR_IC}\\
\mu_{wkb}'(\eta_i) &=&
\frac{1}{\sqrt{2}\beta^{1/4} k^{3/2}}\frac{1}{H \eta_i^2} +i
\frac{\beta^{1/4} k^{3/2}}{\sqrt{2}} H \eta_i
\label{WKBderivative_NLDR_IC}
\end{eqnarray}

To be more accurate the above equations are meaningful only when
\begin{equation}
\Omega^2(\eta)\equiv k^2-\alpha \frac{k^4}{a^2}+\beta \frac{k^6}{a^4}
-\frac{2}{\eta^2}\simeq \beta \frac{k^6}{a^4}
\end{equation}
so one obtains the following constraints on $\beta$:
\begin{equation}
\beta \gg 2 H^2 \frac{a^6}{k^6} \quad,\quad
\beta \gg \frac{a^4}{k^4} \quad,\quad  \beta \gg \alpha \frac{a^2}{k^2} \,.
\end{equation}

The WKB solution is an exact one at the infinite past and thus 
the choice of the adiabatic vacuum is somewhat ``natural''. 
We also recall that all vacua prescriptions are equivalent up to zero-order 
when the WKB approximation holds~\cite{vacua}. 

The choice of a different set of initial conditions at a given $\eta=\eta_1$, 
though, can always be seen as the outcome from a particular (trans-Planckian) 
evolution from a different (and perhaps more consensual) set of initial 
conditions defined at $\eta=\eta_0<\eta_1$. In other words, the choice of 
initial 
conditions is equivalent to the choice of the physics that takes place before 
the moment when they are set. Nevertheless, that choice does not replace the 
discussion on the physics that takes place after that moment, while energies 
above the Planck scale are still at play.

%%%%%%%%%%%%%%%%%%%%%%%%%%%%%%%%%%%%%%%%%%%%%%%
\section{Numerical approach}
\label{num}

In this section we find the correction to the power spectrum, for our 
test scalar field in a de Sitter spacetime with a 
nonlinear d.r., using a fully numerical approach with initial 
conditions given by 
Eqs.~(\ref{WKB_NLDR_IC},\ref{WKBderivative_NLDR_IC}). 
This has been done using a 
C code and the GSL library \cite{gsl}, where we have set $H=0.5 M_{pl}$
and considered a fixed value of $k \gg a(\eta_i) H$, stopping the evolution at
a time $\eta_f$ for which $k \ll a(\eta_f) H$. We have verified that the
result is independent from the value of $k$, namely that, as expected,
the power spectrum is still scale independent.

We write the power spectrum as
\begin{equation}
P_{\phi }(k) = \left(\frac{H}{2\pi }\right)^2
\left[1+ C_{\alpha\beta} \right]\, .
\label{PScorrection}
\end{equation}
The function $C_{\alpha\beta}$ represents the correction due to the nonlinear 
d.r., which obviously depends on the parameters $\alpha$ and $\beta$.

\begin{figure}
\includegraphics[angle=-90,width=0.5\textwidth]{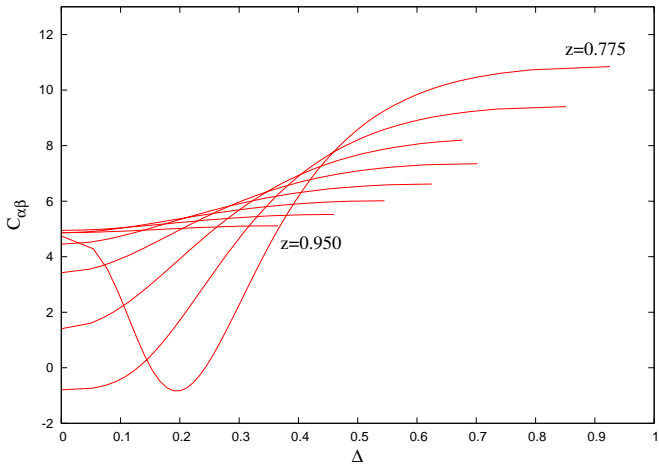}
\caption{Correction $C_{\alpha \beta}$ on the power spectrum, with $z$
varying from 0.775 up to 0.950 with step 0.025, as a function of 
$\Delta$ from a full numerical calculation using the nonlinear 
d.r. (\ref{dr}).}
\label{num_exact_D}
\end{figure}

In Fig.~\ref{num_exact_D} we show the correction $C_{\alpha\beta}$ for 
different values of $z$ (going from $z=0.775$ to $0.950$ with step $0.025$)
in function of $\Delta$. 
We can clearly see 
that $\Delta$ is not the only parameter that plays a role in the calculation 
of the nonlinear
effects on the power spectrum. Indeed, effective potentials with the same 
$\Delta$ have different heights, depths and steepnesses, all of which 
influence effects on the power spectrum. 
Even for small $\Delta$, the correction can be 
large depending on the values of $\alpha$ and $\beta$. 
Therefore, it can hardly be considered a perturbation.

>From Fig.~\ref{num_exact_D}, 
%if we trust to the numerical results, 
we can note that for the smaller value of $z$ one 
obtains a minimum for $C_{\alpha \beta}$ at $\Delta>0$ 
(which corresponds to a finite value of $\beta$ for a fixed $z$).
Besides we have no correction ($C_{\alpha \beta}=0$) for particular values of
$z$ and $\Delta$.  
This means that we could still have no correction in spite of finite 
$\alpha$ and $\beta$. In other words, this nonlinear d.r. yields no 
correction whatsoever to the power spectrum if the parameters happen to be 
around those values. 

%%%%%%%%%%%%%%%%%%%%%%%%%%%%%%%%%%%%%%%%%%%%%%%

\section{Linear approximation}
\label{LA}

Since the WKB factor (\ref{wkb_eq}) is not much smaller than $1$ except for 
$k \gg a H $, one {\it cannot} use the WKB expression (\ref{wkb}) 
for the solution of (\ref{TF_equation_eta2}) (except in the forementioned 
region, of course). Nevertheless, there is indeed an exact solution of 
Eq. (\ref{TF_equation_eta2}) if we approximate the nonlinear
dispersion relation by a straight lines,
as done in Ref.~\cite{rhb}.

\begin{figure}
\includegraphics[angle=-90,width=0.5\textwidth]{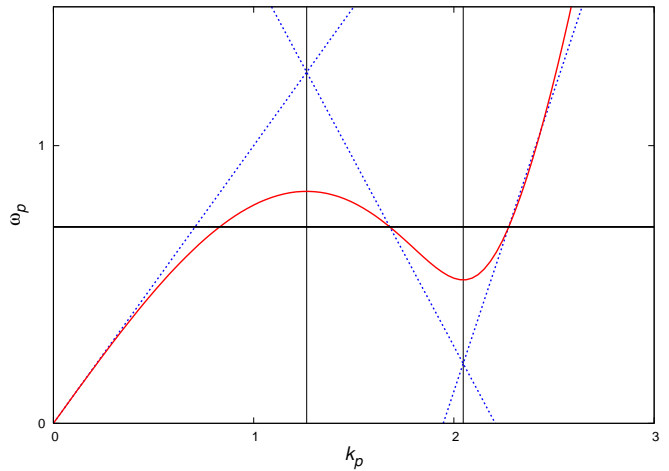}
\caption{Three-piecewise linear approximation to the nonlinear d.r. as 
suggested in Ref.~\cite{rhb}. The horizontal line is $\sqrt{2}H$. Regions 
1,2,3 are defined by the vertical lines and labeled from right to left.
$k_p$ is in units of $M_{pl}$.}
\label{linear}
\end{figure}

If we write $\omega_i (k_p) = A_i k_p + B_i$ for each of the regions $i=1,2,3$ 
(see Fig.~\ref{linear}), the solutions are
\begin{eqnarray}
\mu_i (\eta) &=& c_i W_M\left[\frac{iB_i}{H}, \sqrt{\frac{9}{4}-\gamma_i}, 2i 
A_ik \eta \right] +\nonumber \\
&+& d_i W_W\left[\frac{iB_i}{H}, \sqrt{\frac{9}{4}-\gamma_i}, 2iA_ik\eta 
\right]
\label{mulinear}
\end{eqnarray}
where $\gamma_i\equiv (B_i/H)^2$ and $W_{M,W}(\cdot,\cdot,\cdot)$ are 
Whittaker functions~\cite{abram}. We pick $A_i$ and $B_i$ such that the 
positions of the maximum and minimum of the piecewise d.r. 
coincide with those of the full nonlinear case and that the
first and the second ``turning points" $k_{p 1}$ and $k_{p 2}$  
are the same as well 
(see Fig. \ref{linear}). 
In this way we have the same $\Delta$ of the exact nonlinear d.r.. Of course, 
we pick $A_3=1$ and $B_3=0$, corresponding to the linear d.r. in the 
long-wavelength (small $k_p$) limit.

The coefficients $c_1$, $d_1$ are given by matching the asymptotic 
behavior of the previous equation to the initial conditions yielded by the WKB 
approximation (see below). The former is given by \cite{abram}
\begin{eqnarray}
\mu_1(\eta) &\approx& {\rm e}^{-iA_1k\eta}(2iA_1k\eta)^{+i\frac{B_1}{H}}
\left[c_1 \frac{\Gamma(1+2\nu)}{\Gamma(\frac{1}{2}+\nu+\kappa)} + d_1 \right] 
+ \nonumber \\
&+& {\rm e}^{+iA_1 k\eta} (2iA_1k\eta)^{-i\frac{B_1}{H}}
\left[ c_1 \frac{\Gamma(1+2\nu)}{\Gamma(\frac{1}{2}+\nu-\kappa)}
\right]
\label{mu1asympt}
\end{eqnarray}
where $\nu\equiv \sqrt{9/4-\gamma_1}$ and $\kappa\equiv -2B_1/H$.
Comparing Eq.~(\ref{mu1asympt}) to the expected WKB solution
\begin{eqnarray}
\mu_{wkb}(\eta) &=& \frac{1}{\sqrt{2\omega}}{\rm exp}
\left[-i \int_{\eta_i}^\eta \omega(\eta) d\tau
\right]\nonumber \\
 &=& \frac{1}{\sqrt{2A_1k}}
\left(\frac{\eta}{\eta_i}\right)^ {i\frac{B_1}{H}}{\rm e}^{-i A_1 k 
(\eta-\eta_i)}
\end{eqnarray}
we get
\begin{equation}
\left\{
\begin{array}{l}
c_1 = 0\\
d_1 = \frac{1}{\sqrt{2A_1k}} (2iA_1k\eta_i)^{-iB_1/H}.
\end{array}
\right.
\end{equation}
Note that the expression for $d_1$ reduces to the expected form when $A_1=1$ 
and $B_1=0$, which correspond to the usual d.r..

The amplitude of the growing mode can be calculated as is usually done in 
QM. Since we know the exact solution in each region 
(Eq.~(\ref{mulinear})), all we have to do is to match them and their 
derivatives at the boundaries. In matrix notation, at the boundary between 
regions 1 and 2, we write:
\begin{eqnarray}
&&
\left(
\begin{array}{cc}
W_M^{(1)}(\eta_{12}) & W_W^{(1)}(\eta_{12})\\
 ~ & ~\\
W_M'^{(1)}(\eta_{12}) & W_W'^{(1)}(\eta_{12})
\end{array}
\right)
\left(
\begin{array}{c}
c_1 \\
~ \\
d_1
\end{array}
\right)
= \\
&=&
\left(
\begin{array}{cc}
W_M^{(2)}(\eta_{12}) & W_W^{(2)}(\eta_{12})\\
~ & ~\\
W_M'^{(2)}(\eta_{12}) & W_W'^{(2)}(\eta_{12})
\end{array}
\right)
\left(
\begin{array}{c}
c_2 \\
~\\
d_2
\end{array}
\right)\nonumber
\end{eqnarray}
where the superscripts $(1,2)$ indicate the value of the subscript $i$ in 
Eq.~(\ref{mulinear}), $\eta_{12} \equiv -k_{p 12}/(Hk)$ is the value of $\eta$ 
at that boundary (see Fig.~\ref{linear}) and $(')$ indicates derivative with 
respect to $\eta$. We will write the above equation in a more compact form as
\begin{equation}
{\cal M}_1 \cdot{\cal C}_1 = {\cal M}_2 \cdot{\cal C}_2
\end{equation}
with an obvious notation. Analogously, we can write the matching at the 
boundary of regions 2 and 3 as
\begin{eqnarray}
&&
\left(
\begin{array}{cc}
W_M^{(2)}(\eta_{23}) & W_W^{(2)}(\eta_{23})\\
 ~ & ~\\
W_M'^{(2)}(\eta_{23}) & W_W'^{(2)}(\eta_{23})
\end{array}
\right)
\left(
\begin{array}{c}
c_2 \\
~ \\
d_2
\end{array}
\right)
= \\
&=&
\left(
\begin{array}{cc}
W_M^{(3)}(\eta_{23}) & W_W^{(3)}(\eta_{23})\\
~ & ~\\
W_M'^{(3)}(\eta_{23}) & W_W'^{(3)}(\eta_{23})
\end{array}
\right)
\left(
\begin{array}{c}
c_3 \\
~\\
d_3
\end{array}
\right)\nonumber
\end{eqnarray}
and as
\begin{equation}
{\cal M}_3 \cdot{\cal C}_2 = {\cal M}_4 \cdot{\cal C}_3.
\end{equation}
Such a compact notation allow us to write ${\cal C}_3$ in terms of 
${\cal C}_1$ as
\begin{equation}
{\cal C}_3 = {\cal M}_4^{-1} \cdot {\cal M}_3 \cdot{\cal M}_2^{-1}\cdot 
{\cal M}_1 \cdot{\cal C}_1
\label{C3}
\end{equation}

The coefficient of the growing mode is given by $d_3$ --- the second component 
of ${\cal C}_3$ ---  since it is the coefficient of the divergent Whittaker 
function when 
$\eta \rightarrow 0^-$:
\begin{eqnarray}
W_W\left(-\frac{iB_3}{2H},\frac{\sqrt{1-4\gamma_3}}{2},2iA_3k^2\eta\right) 
\rightarrow \frac{-i}{k\eta}
\end{eqnarray}
while $W_M\rightarrow 0 $ at the same limit. Therefore,
\begin{equation}
\mu_3 (\eta) \rightarrow \frac{-i}{k \eta} d_3
\end{equation}
as $\eta\rightarrow 0^-$. We can thus write the power spectrum as
\begin{eqnarray}
P_{\phi}(k) &=& \frac{k^3}{2\pi^2} {\left|\frac{\mu_3}{a} 
\right|}^2 \nonumber \\
 &\simeq& \frac{H^2 k}{2\pi^2} |d_3|^2,
\end{eqnarray}
in the limit $\eta \rightarrow 0^-$ ($k \ll a H$).
This is, as expected, scale invariant.

\begin{figure}
\includegraphics[angle=-90,width=0.5\textwidth]{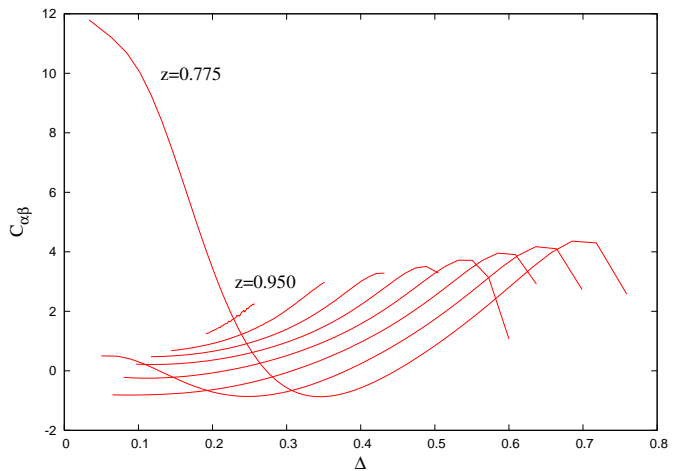}
\caption{Correction to the spectrum using the 3-piecewise approximation, with 
$z$ varying from $0.775$ up to $0.950$ with step $0.025$, 
as a function of $\Delta$.}
\label{3piece-D}
\end{figure}

In Fig.~\ref{3piece-D} we have plotted the correction of the spectrum, due to 
this three-piecewise approximation, 
as a function of $\Delta$. The reason for such large values is that 
the nonlinear d.r. can be qualitatively 
different 
from the linear one ($\omega \sim k$) at large $|\eta|$ even for small 
$\Delta$.

As it is clear, comparing Fig. \ref{3piece-D} to Fig. \ref{num_exact_D}, 
the results obtained with this three-piecewise approximation are pretty 
different from the numerical one obtained with the exact nonlinear d.r..

%%%%%%%%%%%%%%%%%%%%%%%%%%%%%%%%%%%%%%%%%%%%%
\section{Semianalytical approach}

\begin{figure}
\includegraphics[angle=-90,width=0.5\textwidth]{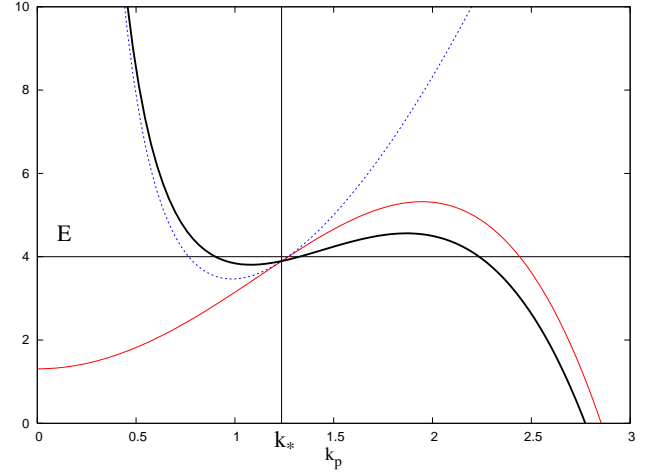}
\caption{Effective potential $V_{\rm eff}(k_p)$ (thick black line) and its 
approximations (thin colored lines) given by $V_{\rm eff}(k_p) = E - 
\Sigma^2(k_p)$ and Eqs.~(\ref{omegas}), for $\bar \alpha=0.1325$, 
$\bar\beta=0.004375$ and $H=0.5 M_{pl}$, as a function of $k_p$ (in $M_{pl}$ 
units). The straight horizontal line is $E\equiv 1/H^2$. The vertical  
line marks the value of $k_*$ (see text for definition).}
\label{veff_fig}
\end{figure}

In this section we use $k_p$ as the independent variable. Since it is 
proportional to $\eta$ ($k_p=k/a=-\eta H k$), there is no particular 
advantage to choose either way, but this variable is more ``physical''\ and 
thus one can rely on her/his physical intuition. For a linear d.r., 
Eq. (\ref{mu}) is then written as
\begin{equation}
 \frac{d^2 \mu_k}{dk_p^2} + \left(\frac{1}{H^2} - \frac{2}{k_p^2} \right) 
\mu_k = 0.
 \label{mu_kp_lin}
 \end{equation}
Using a nonlinear d.r. amounts to the substitution
\begin{equation}
\frac{1}{H^2} \rightarrow  \frac{1}{H^2}\frac{\omega_p^2(k_p)}{k_p^2}
\label{changekp}
\end{equation}
in the previous equation. 
Although there is no exact solution of Eq.~(\ref{mu_kp_lin}) with the 
substitution (\ref{changekp}) 
when one uses the d.r. (\ref{dr}), we still can get a fair analytical 
approximation by writing
\begin{eqnarray}
\Sigma^2(k_p) &\equiv&
\frac{1}{H^2}\frac{\omega_p^2(k_p)}{k_p^2} - \frac{2}{k_p^2}\nonumber\\
 &=&  \frac{\beta}{H^2} k_p^4 -
\frac{\alpha}{H^2}k_p^2 - \frac{2}{k_p^2} + \frac{1}{H^2} \nonumber \\
 &&\hspace{-0.1\textwidth}\approx \left\{
        \begin{array}{lc}
        \Sigma_1^2(k_p) \equiv \frac{\beta}{H^2} k_p^4 -
\frac{\alpha}{H^2}k_p^2 +\frac{1}{H^2} - c & , k_p\geq k_* \\
 ~ & ~ \\
        \Sigma_2^2(k_p) \equiv -\frac{\alpha}{H^2}k_p^2 + \frac{1}{H^2} -
\frac{2}{k_p^2} - d & , k_p\leq k_*
        \end{array}
\right.
\label{approx}
\end{eqnarray}
where $c$, $d$ and the matching point $k_*$ are defined by requiring that
\begin{equation}
\left\{
\begin{array}{l}
\Sigma_1^2(k_*) =\Sigma_2^2(k_*)=\Sigma^2(k_*) \\
~ \\
\left.\frac{d\Sigma_1^2}{dk}\right|_{k_*}=\left.\frac{d\Sigma_2^2}
{dk} \right|_{k_*}
\end{array}
\right.,
\label{omegas}
\end{equation}
which correspond to $k_*=(H^2/\beta)^{1/6}$, $d=-(\beta/H^2)^{1/3}$ and
$c= 2(\beta/H^2)^{1/3}$. The approximated expressions have the same limiting 
behaviors (both at $k_p \rightarrow 0$ and $k_p \rightarrow \infty$) than the 
original one. That feature provides another piece of information on 
the dependence of the outcome on the details of the nonlinear d.r. for 
intermediate $k_p$, as opposed to its small- and large-scale limits. See 
Fig.~\ref{veff_fig} for a qualitative comparison.

The calculation itself is carried out by the usual approach in QM, as in the 
previous section. The requirements that the ``wave function" and its first 
derivative are continuous at $k_*$ are easily accomplished since the solutions 
of the equation of motion (\ref{mu_kp_lin},\ref{changekp}) with the 
approximations (\ref{approx}) are known analytically. For $k>k_*$ we find:
\begin{equation}
\frac{d^2 f_1(k_p)}{dk_p^2} - \Sigma_1^2(k_p) f_1(k_p) = 0,
\end{equation}
\begin{eqnarray}
f_1(k_p) &=& c_1 \exp\left[+\frac{i}{H}\left(-\frac{\alpha}{2\sqrt{\beta}} k_p
+\frac{\sqrt{\beta}}{3}k_p^3\right)\right]  Ht_1(k_p) + \nonumber\\
&&\hspace{-0.7cm} +d_1 \exp\left[-\frac{i}{H}\left(
-\frac{\alpha}{2\sqrt{\beta}} k_p+\frac{\sqrt{\beta}}{3}k_p^3\right)\right]
Ht_2(k_p)
\label{f1}
\end{eqnarray}
where $Ht_1(k_p) \equiv HeunT(A,0,B,+i\zeta)$ and $Ht_2(k_p) \equiv
HeunT(A,0,B,-i\zeta)$ are triconfluent Heun functions~\cite{heun}, with
\begin{eqnarray}
A&\equiv&\left(\frac{3}{2H^2\sqrt{\beta}}\right)^{2/3}\left(-1+cH^2+
\frac{\alpha^2}{4\beta}\right),
\\
B &\equiv& \left(\frac{3}{2H^2\beta^2}\right)^{1/3}\alpha\\
\zeta &\equiv& \left(3\sqrt{2}\beta^{1/4}H\right)^{1/3}k_p.
\end{eqnarray}
 We can constrain the coefficients $(c_1, d_1)$ in Eq.~(\ref{f1}) by 
comparing the asymptotic behavior of the above equation to the expected WKB 
one, Eq.~(\ref{WKB_NLDR}). The asymptotic limit of interest here is taken 
along a Stokes line ($arg(i\zeta)=\pi/2$) of the triconfluent Heun function, 
which means that there are two equally dominant terms in the asymptotic 
expansion:
\begin{equation}
HeunT(A,0,B,i\zeta) \sim
\zeta^{-1}[a+b \exp(-i\zeta^3)]
\label{limit}
\end{equation}
as $\zeta\rightarrow +\infty$, where $a$ and $b$ are constants. Since neither 
one is dominant over the other, both must be taken into account. 
The forementioned comparison yields
\begin{equation}
\left\{
\begin{array}{l}
d_1 b - c_1 a = 0\\
- c_1 b + d_1 a = C
\end{array}
\right.
\label{sys}
\end{equation}
where $C\equiv i/[(3H\sqrt{2}\beta^{1/4})^{1/3}\sqrt{k}]$. There are only 2 
equations for 4 unknowns, which should be expected from the lack of a dominant 
behavior in the asymptotic expansion along a Stokes line, as mentioned above. 
One could choose whichever 2 of the above parameters (say, $a$ and $b$) to be 
specified by comparing the final result to the outcome of the numerical 
calculation using the approximated effective potential. Formally, that 
determination is supposed to be done at every value of $\alpha$ and $\beta$, 
which would render our semianalytical approach useless.  Nevertheless, we
have found that writing $b$ and $d_1$ in terms of $a$ and $c_1$ yields a 
qualitatively good behavior for different values of $z$, i.e, for different 
pairs $\{\alpha,\beta\}$ when compared to the numerical calculation 
(see Fig.~\ref{alls} below). Such procedure yields
\begin{eqnarray}
b &=& \frac{c_1}{d_1} a\\
d_1 &=& \frac{-C \pm \sqrt{C^2 + 4a^2c_1^2}}{2a}\label{d1}
\end{eqnarray}
The sign in the above equation was again numerically determined to be the 
lower one (see below). Following this line of reasoning, we used $a=0.1$ 
and $c_1=-1.8$ for all $z$.

The last step of the semianalytical procedure is the evolution in the second 
region, when $k<k_*$:
\begin{equation}
\frac{d^2 f_2(k_p)}{dk_p^2} - \Sigma_2^2(k_p) f_1(k_p) = 0,
\end{equation}
whose exact solution is
\begin{eqnarray}
f_2(k_p) &=& c_2 \frac{1}{\sqrt{k_p}} W_M\left(D,\frac{3}{4},
\frac{\sqrt{\alpha}k_p^2}{H}\right)+\nonumber \\
&+& d_2 \frac{1}{\sqrt{k_p}} W_W\left(D,\frac{3}{4},
\frac{\sqrt{\alpha}k_p^2}{H}\right)
\end{eqnarray}
where $D \equiv (1-dH^2)/(4H\sqrt{\alpha})$.

The coefficient $d_2$ determines the amplitude of the perturbations since the
function $W_W(k_p)$ is the growing solution:
\begin{equation}
\lim_{k_p\rightarrow 0} \frac{W_W(k_p)}{\sqrt{k_p}} =
{\frac {\sqrt {\pi}H^{1/4}}{2 \alpha^{1/8} \Gamma (\xi) } }\frac{1}{k_p}
\end{equation}
where $\xi \equiv \left(
\frac{5}{4}-\frac{1-dH^2 }{4H\sqrt{\alpha}}\right)$.
The above expression allows us to write the spectrum as
\begin{eqnarray}
P_{\phi}(k) &\equiv& \frac{k^3}{2\pi^2} \left|
\frac{f_2}{a} \right|^2 \\
 &\simeq& \frac{k\sqrt{H}}{8\pi \alpha^{1/4} \Gamma^2 (\xi)} |d_2|^2 \\
 &=& \left(\frac{H}{2\pi}\right)^2 (1 + C_{\alpha\beta})\,,
\label{spec}
\end{eqnarray}
in the superhorizon limit $k_p \ll H$.
The coefficient $d_2$ is determined by the forementioned procedure, requiring 
that the function and its derivative are continuous at $k=k_*$. In matrix 
notation, it can be written as
 \begin{eqnarray}
 &&\left(
 \begin{array}{cc}
 g_1(k_*) & h_1(k_*)\\
 ~&~\\
 \left(\frac{dg_1(k_p)}{dk_p}\right)_{k_p=k_*} & 
\left(\frac{dh_1(k_p)}{dk_p}\right)_{k_p=k_*}
 \end{array}
 \right)
 \left(
 \begin{array}{c}
 c_1\\
 ~\\
 d_1
 \end{array}
 \right)
 =\nonumber\\
 &&=
\left(
 \begin{array}{cc}
 g_2(k_*) &  h_2(k_*)\\
 ~&~\\
 \left(\frac{dg_2(k_p)}{dk_p}\right)_{k_p=k_*}
 &
 \left(\frac{dh_2(k_p)}{dk_p}\right)_{k_p=k_*} \\
 \end{array}
 \right)
 \left(
 \begin{array}{c}
 c_2 \\
~\\
 d_2
 \end{array}
 \right)
 \end{eqnarray}

where
\begin{eqnarray}
\!\!\!g_1(k_p) &\equiv& \exp\left[\frac{+i}{H}\left(\frac{-\alpha}{2\sqrt{\beta}} 
k_p+\frac{\sqrt{\beta}}{3}k_p^3\right)\right]  Ht_1(k_p)\\
\!\!\!h_1(k_p) &\equiv&
 \exp\left[\frac{-i}{H}\left(\frac{-\alpha}{2\sqrt{\beta}} k_p+
\frac{\sqrt{\beta}}{3}k_p^3\right)\right] Ht_2(k_p)\\
\!\!\!g_2(k_p) &\equiv& \frac{1}{\sqrt{k_p}} W_M\left(D,\frac{3}{4},
\frac{\sqrt{\alpha}k_p^2}{H} \right)\\
\!\!\!h_2(k_p) &\equiv&
 \frac{1}{\sqrt{k_p}} W_W\left(D,\frac{3}{4},\frac{\sqrt{\alpha}k_p^2}{H} 
\right)
\end{eqnarray}
which can be written in a more compact form, as in the previous section, as
\begin{equation}
{\cal M}_1 \cdot {\cal C}_1 = {\cal M}_2 \cdot {\cal C}_2.
\end{equation}
As before, one can invert such equation and write
\begin{equation}
{\cal C}_2 = {\cal M}_2^{-1} \cdot {\cal M}_1 \cdot {\cal C}_1,
\end{equation}
which shows once more that $d_2$ is a linear combination of $c_1$ and $d_1$.
Since we have fixed $c_1$ from the beginning, we are not able to find 
analytically the dependence of $d_2$ on $k$ and, therefore, to say if the 
spectrum is scale invariant, as it should be. Nevertheless, we approach the 
same problem numerically and we verify that the problem exhibits such 
invariance.

In order to be able to measure the acuracy of the semianalytical approach, 
we have used the approximation (\ref{approx}) and numerically evolved the 
initial conditions (\ref{WKB_NLDR_IC}, \ref{WKBderivative_NLDR_IC}). 
The evolution was split in two pieces: for 
$k>k_*$ and $k<k_*$. Such calculation, besides fixing the sign in 
Eq.~(\ref{d1}), also yields the ``best"  (i.e., more robust with respect to
changes in 
$\alpha$ and $\beta$) values of $c_1=-1.8$ and $a=0.1$.

In Fig.~\ref{alls} we show the function $C_{\alpha\beta}$ for different 
values of $z$ as a function of $\Delta$, as given by the numerical evolution 
using 
the approximated expressions for the effective potential and as given by the 
semianalytical approach with the same approximation. 
We can see that the correction depends also on 
$z$, showing that $\Delta$ is not the only variable to work with.

The semianalytical solution is not expected to work well for small 
$\bar\beta$, which means, for a fixed $z$, small $\Delta$. 
For that, we should have taken into account the next-to-leading 
order terms in the asymptotic expansion of the Heun functions, 
Eq.~(\ref{limit}). That would have, however, introduced new parameters that 
would 
have had to be fixed by comparing to the numerical solution once more. Since we
see no advantage in having a large number of such parameters, we did not do so.

We note that we reproduce a minimum $C_{\alpha\beta}$ at a finite $\Delta$ as 
in the previous section. 
This approximation yields a correction $C_{\alpha \beta}$ with a good 
qualitative
behavior as compared to Fig. \ref{num_exact_D}, but a large difference in the
magnitude of the effect.

\begin{figure}
\includegraphics[angle=-90,width=0.5\textwidth]{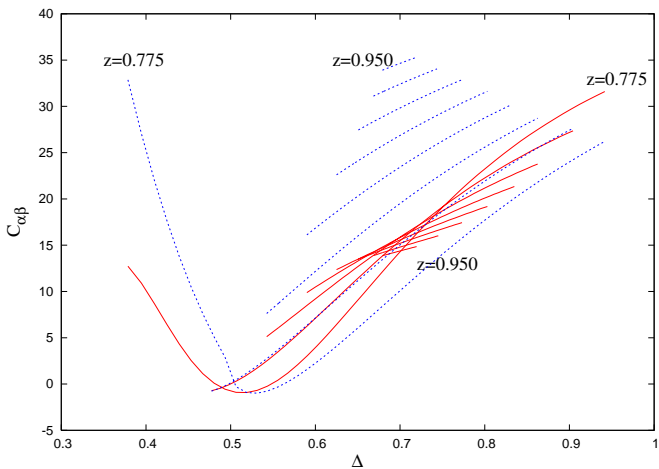}
\caption{Correction $C_{\alpha\beta}$ as a function of $\Delta$ when $z$ 
varies from $0.775$ to $0.950$ (as shown by the labels) for the
semianalytical approach (dashed blue lines) and  for the numerical 
evolution with 
approximated effective potential (\ref{approx}) (full red lines).}
\label{alls}
\end{figure}

%%%%%%%%%%%%%%%%%%%%%%%%%%%%%%%%%%%%%%%%%%%%%%
\section{Power-law inflation}
\label{power}

A numerical approach can also be applied to find the correction 
to the power spectrum in a realistic model of inflation for the nonlinear 
dispersion relation under consideration. 
In this section we restrict ourselves to the case of power-law inflation 
\cite{LMLS} where, in proper time,  $a(t)\sim t^p$,
with $p>1$. This expansion is generated by an exponential potential
\begin{equation}
V(\phi)=V_0 \exp \left[-\frac{\lambda}{M_{pl}}(\phi-\phi_i)\right]
\end{equation}
with $V_0=\frac{M_{pl}^2}{t_i^2} p(3p-1)$ and
$\lambda=\left(\frac{2}{p}\right)^{1/2}$.
The scale factor and the homogeneus solution of the scalar field are given by
\begin{eqnarray}
a(t) &=&\left(\frac{t}{t_i}\right)^p ,\\
\phi(t) &=&\phi_i+M_{pl}(2 p)^{1/2} \log \frac{t}{t_i}
\end{eqnarray}
respectively, and the slow-roll parameters by
\begin{equation}
\begin{array}{lcr}
\epsilon_1 \equiv \frac{M _{pl}^2}{2}\left(\frac{V_\phi}{V}\right)^2=\frac{1}{p}
&
,&
\epsilon_2 \equiv M_{pl}^2 \frac{V_{\phi\phi}}{V}=\frac{2}{p},
\end{array}
\end{equation}
where $V_{\phi}\equiv d V/d\phi$ and so on. 
In conformal time $\eta$, the scale factor and the Hubble parameter become
\begin{equation}
\begin{array}{lcr}
a(\eta)=\left(\frac{\eta}{\eta_i}\right)^{\frac{p}{1-p}}&
,&
H(\eta)=-\frac{p}{p-1}\left(\frac{\eta}{\eta_i}\right)^{\frac{p}{p-1}}
\frac{1}{\eta}
\end{array}
\label{aeta}
\end{equation}
with $\eta_i=t_i/(1-p)$. 
The equation of motion for the Mukhanov variable, Eq.(\ref{EqMuk}), becomes 
(using $\eta$ as the independent variable)
\begin{equation}
\mu_{s k}''+ \left( k^2-\frac{a''}{a}\right) \mu_{s k}=0 \,
\label{TF_equation_eta_powerlaw}
\end{equation}
and the canonically normalized solution, associated with the adiabatic 
vaccum for $k \gg a H$, is given by (see, for example, \cite{Marozzi})
\begin{equation}
\mu_k=\left(-\frac{\pi \eta}{4}\right)^{1/2} H_\nu^{(1)}
    (-k \eta) \,,
\end{equation}
with $\nu=3/2+1/(p-1)$.
So, in the long-wavelength limit, we obtain the following scale dependent
power spectrum
\begin{equation}
P_{Q}(k)=\frac{1}{(2 \pi)^2} \left(-\frac{2}{\eta_i}\right)^{
\frac{2 p}{p-1}} \frac{1}{\pi} \Gamma^2(\nu) k^{-\frac{2}{p-1}}
\label{PSpowerlaw}
\end{equation}
with a spectral index $n_s=1-\frac{2}{p-1}$.
Using this result, namely the dependence on k of the power spectrum,
the range $p<60$ is disfavored at $2\sigma$ by the observation
(see, for example, \cite{ObsPL}).

Now we want to see if the introduction of the particular nonlinear d.r. 
given in Eq. (\ref{dr}) changes the dependence of the power spectrum on $k$.
Let us set our background: Eq.~(\ref{TF_equation_eta_powerlaw}) can be
written, using Eq.~(\ref{aeta}), as
\begin{equation}
\mu_k''+ \left[ k^2-\frac{2 p^2-p}{(1-p)^2} \frac{1}{\eta^2}\right] \mu_k=0 \,.
\label{TF_equation_eta_powerlaw_in_eta}
\end{equation}
For $p \gg 1$, this can be approximated as
% %
\begin{equation}
\mu_k''+ \left[ k^2-2 a^2 H^2\right] \mu_k=0 \,,
\label{TF_equation_eta_powerlaw_approximate}
\end{equation}
so, in this limit and following the same reasoning as before, 
to obtain a positive nonlinear d.r. that gives three real
distinct classical turning-points ($\omega_p^2=2 H^2$) 
at the initial time \footnote{The Hubble 
factor is time-dependent but for $p \gg 1$ it becomes almost constant 
(see Eq.(\ref{aeta})).} we require that 
\begin{equation}
3/4<z<1 \,\,\,\,\,\,\,,\,\,\,\,\,\,\,
\frac{g(z)}{H(\eta_i)^2}<\alpha<\frac{f(z)}{H(\eta_i)^2}\,.
\end{equation}
with $g(z)$ and $f(z)$ given in Eqs. (\ref{grange},\ref{frange}).

The new equation of motion will be given by
\begin{equation}
\mu_k''+ \left[ a^2\left(\frac{k^2}{a^2}-\alpha \frac{k^4}{a^4}
+\beta \frac{k^6}{a^6}\right)
-\frac{2 p^2-p}{(1-p)^2} \frac{1}{\eta^2}\right] \mu_k=0 \,
\label{TF_equation_eta_powerlaw_in_eta_non_linear}
\end{equation}
and, repeating the calculation in  Sec. \ref{ic} with the new background,
one obtains the following initial conditions
\begin{eqnarray}
\mu_{wkb}(\eta_i) &=& \frac{1}{\sqrt{2}\beta^{1/4} k^{3/2}}
\label{WKB_NLDR_IC_powerlaw}\\
\mu_{wkb}'(\eta_i) &=&
-\frac{1}{\sqrt{2}\beta^{1/4} k^{3/2}} \frac{p}{p-1}\frac{1}{\eta_i} -i
\frac{\beta^{1/4}}{\sqrt{2}} k^{3/2}
\label{WKBderivative_NLDR_IC_powerlaw}
\end{eqnarray}

\begin{figure}
\includegraphics[angle=-90,width=0.5\textwidth]{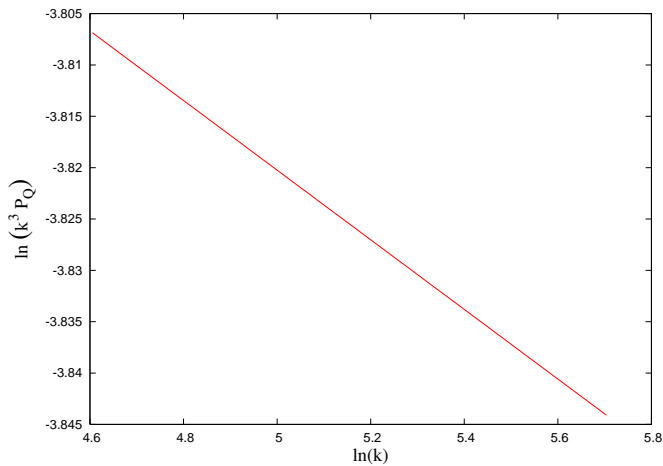}
\caption{Logarithm of the power spectrum in a power-law model of inflation 
in function of $\ln(k)$ for a linear d.r.. $k$ is in units of $M_{pl}$.}
\label{PSpowerlaw_fig}
\end{figure}

\begin{figure}
\includegraphics[angle=-90,width=0.5\textwidth]{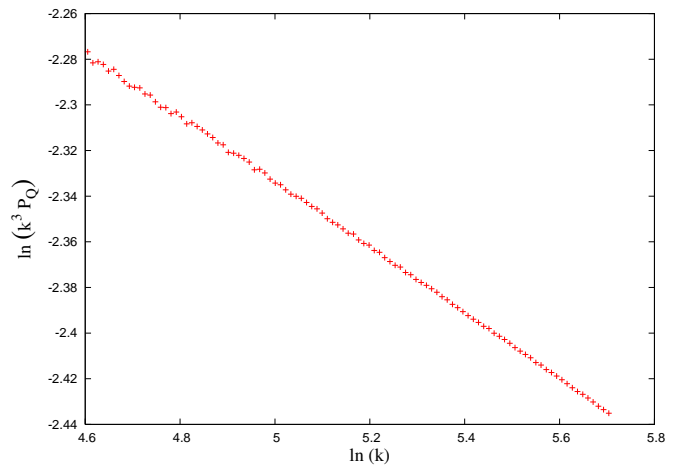}
\caption{Logarithm of the power spectrum in a power-law model of inflation 
using the nonlinear d.r. (\ref{mu}) in function of $\ln(k)$.
$k$ is in units of $M_{pl}$.}
\label{PSpowerlawNonlinear}
\end{figure}

We now proceed with the numerical analysis taking $t_i=(p-1)/M_{pl}$ (and thus
$\eta_i=-1/M_{pl}$). The restriction to apply the adiabatic initial condition
($-k \eta_i \gg 1$) becomes equivalent to $k \gg M_{pl}$; therefore, 
we are indeed in the 
energy scale of TPP. Besides, with those initial conditions and in the limit
$p \gg 1$, we have 
$H(\eta_i)\simeq M_{pl}$ and the new d.r. 
becomes very different from the linear one only for $k={\cal O}(M_{pl})$.
For the numerical analysis we take $z=0.80$, 
$\alpha=\frac{f(0.80)+g(0.80)}{2 H(\eta_i)^2}$ and the limiting value $p=60$, 
making the analysis
for $k$ between $100 M_{pl}$ and $300 M_{pl}$ and going from $\eta_i$ to a
time for which $k \ll a H$.
We give in Fig. \ref{PSpowerlaw_fig} the logarithm of the power spectrum 
calculated for the standard linear case in function of $\ln(k)$, 
while in  Fig. \ref{PSpowerlawNonlinear} the logarithm of the power spectrum 
calculated for the nonlinear d.r. case in function of  $\ln(k)$ for many 
different points inside the aforementioned range of $k$.
As one can see from this plot, the power spectrum for the nonlinear case 
still has a power-law dependence on  $k$, but this d.r. produces a change 
both in the normalization factor and in the spectral index when compared to 
the linear case.
Fitting the data in Fig.~\ref{PSpowerlawNonlinear} we obtain a spectral 
index smaller than that of the linear case: $0.857$ instead of $0.966$.
The new spectral index would be almost the same as that of the linear 
case if we 
have taken $p \sim 15$. One obtains similar results starting from
different values of $p$ or with different values for the parameters 
$\alpha$ and $\beta$. Therefore, the introduction of the nonlinear d.r. 
gives a stronger dependence of the power spectrum on $k$ and, as a 
consequence, a possible different range of disfavored values for $p$.

A different approach to this problem in a power-law model is shown in 
\cite{Tronconi}. The authors use the minimum-uncertainty principle to fix the initial conditions and also find corrections to the power spectrum. We also would like to underline that our results are not in disagreement with Ref.~\cite{StaroPL}. Indeed, we consider a regime for which $H$ is of the order of $M_{pl}$ while the results of \cite{StaroPL} assume $H\ll M_{pl}$.

\begin{figure}
\includegraphics[width=0.5\textwidth]{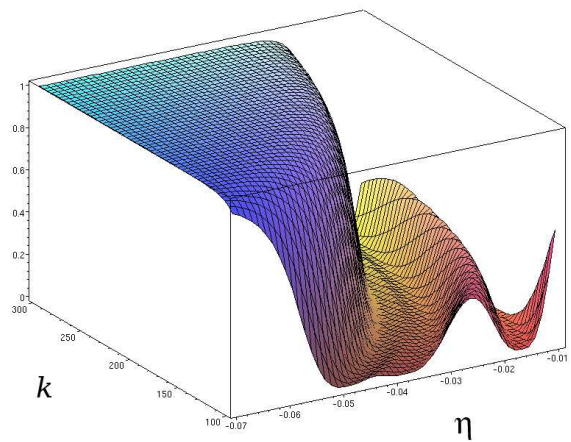}
\caption{Plot of the function $1/(1+W)$ in terms of the wavenumber $k$ (in units of $M_{pl}$) and the conformal time $\eta$ using the same parameters as in Fig.~\ref{PSpowerlawNonlinear} (see text for the precise values). One can clearly see that the WKB approximation holds ($W\ll 1$) only for early times.}
\label{wkb3d}
\end{figure}

Using the same parameters as in Fig.~\ref{PSpowerlawNonlinear}, we have plotted in  Fig.~\ref{wkb3d} the behavior of $1/(1+W)$, where $W$ is defined in Eq.~(\ref{wkb_eq}). One can easily see that the WKB approximation does not hold at small absolute values of the conformal time, which stresses our very point\footnote{Although WKB is not a good approximation, we have calculated the spectral index with the nonlinear d.r. using, in a strict and forward way, the WKB solutions (\ref{wkb}) and the same parameters as before, for the sake of comparison. The spectrum turns out to be strongly (blue) shifted from the correct (numerical) value: $n_s^{\rm wkb} = 1.22$.}.

%%%%%%%%%%%%%%%%%%%%%%%%%%%%%%%%%%%%%%%%%%%%%%
\section{Conclusions}
\label{Conc}

In this paper we have shown that the nonlinear d.r. (\ref{dr}) presents 
nonlinear results, i.e, which are not shown in the perturbative approaches 
used in previous papers in the subject. 
In particular, the correction $C_{\alpha\beta}$ can be almost as large as 10
depending on the value of the parameters $\alpha$, $\beta$. Therefore, it can 
hardly be considered a perturbation. 

Our results agree, in part, with the literature on the effects on the power 
spectrum obtained using different approachs to mimic TPP.
In fact, also using different approachs, the effects are always of the order
of $H/M_{pl}$ (see \cite{rhb} 
and references cited therein) which in our case is ${\cal O} (1)$.

We have compared previously used approximations and introduced a new one, 
which allowed a semianalytical calculation. 

The fundamental point behind such a large correction factor lies in the break 
of the WKB approximation at early (conformal) times. If the d.r. at this 
moment is 
highly nonlinear, as it happens here, the corrections are bound to be 
large \cite{dan}. 

However, from Fig. \ref{num_exact_D} we can see that for particular value of 
$z$ and $\Delta$ we obtain no correction to the power spectrum.
Namely, that we could still have no correction in spite of finite $\alpha$ and 
$\beta$. In other words, this nonlinear d.r. yields no correction whatsoever 
to the power spectrum if the parameters happen to be around those values.

In the last section we have  used a power-law model of inflation as the 
background in order to understand what features were particular to the de 
Sitter (exponential) expansion. 
It is clear that one would always get a scale-free spectrum in 
a de Sitter background, but what would happen to the spectral index $n_s$ in 
a different one? We have shown that this particular d.r. (\ref{dr}) yields 
a even smaller $n_s$ as compared to power-law inflation with a linear d.r., 
and therefore, it is strongly disfavored by observational results.  
On the other hand, this result suggests that another nonlinear d.r. 
may yield also a different correction (perhaps in the opposite direction) to 
the value of $n_s$ in a power-law background.  

We have not addressed the question on back-reactions which may play an 
important role, since the corrections to the power spectrum are not 
perturbative ones. Such implications are beyond the scope of this work and 
deserve a separate study, which will be published elsewhere.

%%%%%%%%%%%%%%%%%%%%%%%%%%%%%%%%%%%%%%%%%%%%%%%%%%%%%%%%%%%%%%%%%%%%

{\bf Acknowledgments:} We would like to thank R. Ansari, A. Sarkar and 
M. Schulze for useful discussions and 
Professor R.H. Brandenberger for useful correspondence. 
S.E.J. thanks Professor Duval for her help on the Heun functions 
and acknowledges financial support 
from ICTP and from CNPq.

%%%%%%%%%%%%%%%%%%%%


\begin{thebibliography}{99}

\bibitem{rhb99} R.H. Brandenberger, {\tt hep-ph/9910410}.
 %%CITATION = HEP-PH/9910410;%%

\bibitem{first} J. Martin and R.H. Brandenberger, Phys. Rev. {\bf D 63},
123501 (2001); R.H. Brandenberger and J. Martin, Mod. Phys. Lett. A
{\bf 16}, 999 (2001).
%%CITATION = PHRVA,D63,123501;%%
%%CITATION = MPLAE,A16,999;%%

\bibitem{w_phys} M. Lemoine, M. Lubo, J. Martin and J.-P. Uzan, Phys. Rev.
{\bf D 65}, 023510 (2001).
%%CITATION = PHRVA,D65,023510;%%

\bibitem{rhb} J. Martin and R. Brandenberger, Phys. Rev. {\bf D 68},
063513 (2003).
%%CITATION = PHRVA,D68,063513;%%

\bibitem{Grish} L. P. Grishchuk, Zh. Eksp. Teor. Fiz. {\bf 67}, 825 (1974)
[Sov. Phys. JETP {\bf 40}, 409 (1975)].
%%CITATION = ZETFA,67,825;%%

\bibitem{Muk} V. F. Mukhanov, Zh. Eksp. Teor. Fiz. {\bf 94},No. 7, 1 (1988)
[Sov. Phys. JETP {\bf 67}, 1297 (1988)].
%%CITATION = ZETFA,94N7,1;%%

\bibitem{merz} Eugen Merzbacher, ``Quantum Mechanics'' (John Wiley \& Sons, 1970),
$3^{\rm rd}$ edition.

\bibitem{wkb} J. Martin and D. J. Schwarz, Phys. Rev. {\bf D 67}, 083512
(2003)
%%CITATION = PHRVA,D67,083512;%%

\bibitem{unruh} W.G. Unruh, Phys. Rev {\bf D 51}, 2827 (1995).
%%CITATION = PHRVA,D51,2827;%%

\bibitem{cj} S. Corley and T. Jacobson, Phys. Rev. {\bf D 54}, 1568 (1996).
%%CITATION = PHRVA,D54,1568;%%

\bibitem{kg} S. Cremonini, Phys. Rev. {\bf D 68}, 063514 (2003); J.
Kowalski-Glikman, Phys. Lett. {\bf B 499}, 1 (2001); H.-C. Kim,
 J. H. Yee and C. Rim, Phys. Rev. {\bf D 72}, 103523 (2005).
%%CITATION = PHRVA,D68,063514;%%
%%CITATION = PHLTA,B499,1;%%
%%CITATION = PHRVA,D72,103523;%%

\bibitem{abrik} A.A. Abrikosov, ``Methods of Quantum Field Theory in
Statistical Physics'' (Dover Pub., 1975);  T. Jacobson, Phys. Rev. {\bf D 44},
1731 (1991).
%%CITATION = PHRVA,D44,1731;%%

\bibitem{myers} Robert C. Myers and M. Pospelov, Phys. Rev. Lett. {\bf 90},
211601 (2003).
%%CITATION = PRLTA,90,211601;%%

\bibitem{rhb2} R.H. Brandenberger and J. Martin, Phys. Rev. {\bf D 71},
023504 (2005).
%%CITATION = PHRVA,D71,023504;%%

\bibitem{vacua} U. H. Danielsson, Phys. Rev. {\bf D 66}, 023511 (2002).
%%CITATION = PHRVA,D66,023511;%%

\bibitem{gsl}{\tt http://www.gnu.org/software/gsl}

\bibitem{abram} M. Abramowitz and I.A. Stegun, ``Handbook of mathematical 
functions with formulas, graphs, and mathematical tables" (Washington, D.C.: 
National Bureau of Standards, 1964).
 
\bibitem{heun} A. Ronveaux (Ed.), ``Heun's differential equations"\ 
(Oxford University Press, 1995).

\bibitem{LMLS} F. Lucchin and S. Matarrese, Phys. Rev. {\bf D 32}, 1316 (1985);
D. Lyth and E. Stewart, Phys. Lett. {\bf B 274}, 168 (1992).
%%CITATION = PHRVA,D32,1316;%%
%%CITATION = PHLTA,B274,168;%%

\bibitem{Marozzi} G. Marozzi, Phys. Rev. {\bf D 76}, 043504 (2007).
%%CITATION = PHRVA,D76,043504;%%

\bibitem{ObsPL} F. Finelli, M. Rianna and N. Mandolesi,
J. Cosmol. Astropart. Phys. 12 (2006) 006.
%%CITATION = JCAPA,0612,006;%%

\bibitem{Tronconi} G.L. Alberghi, R. Casadio and A. Tronconi, Phys. Lett. 
{\bf B 579}, 1 (2004).
%%CITATION = PHLTA,B579,1;%%


\bibitem{StaroPL} A. A. Starobinsky,
JETP Lett. {\bf 73}, 371 (2001).
%%CITATION = JTPLA,73,371;%%

\bibitem{dan} J.C. Niemeyer and R. Parentani, Phys. Rev. {\bf D 64},
101301(R) (2001).  
%%CITATION = PHRVA,D64,101301;%%

\end{thebibliography}
\end{document}